# Towards a vision foundation model for comprehensive assessment of Cardiac MRI

Athira J Jacob[a,b,*], Indraneel Borgohain[a], Teodora Chitiboi[a,c], Puneet Sharma[a], Dorin Comaniciu[a], Daniel Rueckert[b,d]

[a]*Digital Technology and Innovation, Siemens Healthineers, Princeton, NJ, USA*
[b]*AI in Healthcare and Medicine, Klinikum rechts der Isar, Technical University of Munich, Germany*
[c]*Digital Technology and Innovation, Siemens Healthineers AG, Hamburg, Germany*
[d]*Department of Computing, Imperial College London, UK*

ABSTRACT

Cardiac magnetic resonance imaging (CMR), considered the gold standard for noninvasive cardiac assessment, is a diverse and complex modality requiring a wide variety of image processing tasks for comprehensive assessment of cardiac morphology and function. Advances in deep learning have enabled the development of state-of-the-art (SoTA) models for these tasks. However, model training is challenging due to data and label scarcity, especially in the less common imaging sequences. Moreover, each model is often trained for a specific task, with no connection between related tasks. In this work, we introduce a vision foundation model trained for CMR assessment, that is trained in a self-supervised fashion on 36 million CMR images. We then finetune the model in supervised way for 9 clinical tasks typical to a CMR workflow, across classification, segmentation, landmark localization, and pathology detection. We demonstrate improved accuracy and robustness across all tasks, over a range of available labeled dataset sizes. We also demonstrate improved few-shot learning with fewer labeled samples, a common challenge in medical image analyses. We achieve an out-of-box performance comparable to SoTA for most clinical tasks. The proposed method thus presents a resource-efficient, unified framework for CMR assessment, with the potential to accelerate the development of deep learning-based solutions for image analysis tasks, even with few annotated data available.

## 1. Introduction

Cardiac Magnetic Resonance (CMR) Imaging is considered the reference standard for non-invasively assessing cardiac structure, function, and viability. However, a typical CMR study is complex and consists of different sequences, requiring a wide variety of image analysis tasks. For example, segmentation of the left and right ventricles (LV, RV) and left and right atria (LA, RA) on cine images throughout the cardiac cycle are required for assessing ejection fraction, and other functional metrics. Segmentation of the left ventricle myocardium in delayed gadolinium enhancement CMR (LGE) and in tissue mapping images (T1 mapping pre- or post- contrast, T2 mapping, ECV etc.) is required for myocardial tissue characterization. In addition, the essential step of pre-filtering and selecting the appropriate cardiac sequences and views for each of these analyses, is also required. Reporting of myocardial analyses is standardized clinically through the AHA segment model (Cerqueira et al., 2002), which additionally requires the detection of key points (landmarks) in the cardiac region. Integration of information from multiple techniques is required for diagnosing cardiac pathologies. The insights gathered from such an analysis could also influence study acquisition, where the measured cardiac parameters influence the scan plan during the acquisition process. However, assessment of CMR images requires high clinical expertise and is time-consuming, due to variability in the images from differing scan protocols, scanners, normal and pathological anatomical variations, image artifacts, and the lack of quantitative standards. Several studies have explored each of these tasks individually or together, and deep learning (DL) techniques have emerged as the state-of-the-art (SoTA) across the various tasks. However, many of these studies are on very small cohorts or are limited to cine MRI on relatively healthy subjects. An additional consequence of the fragmented landscape is that, in general, each of the cardiac image analysis tasks, though highly related, are treated independently with no in-formation sharing between the trained DL networks.

---

[*]Corresponding author

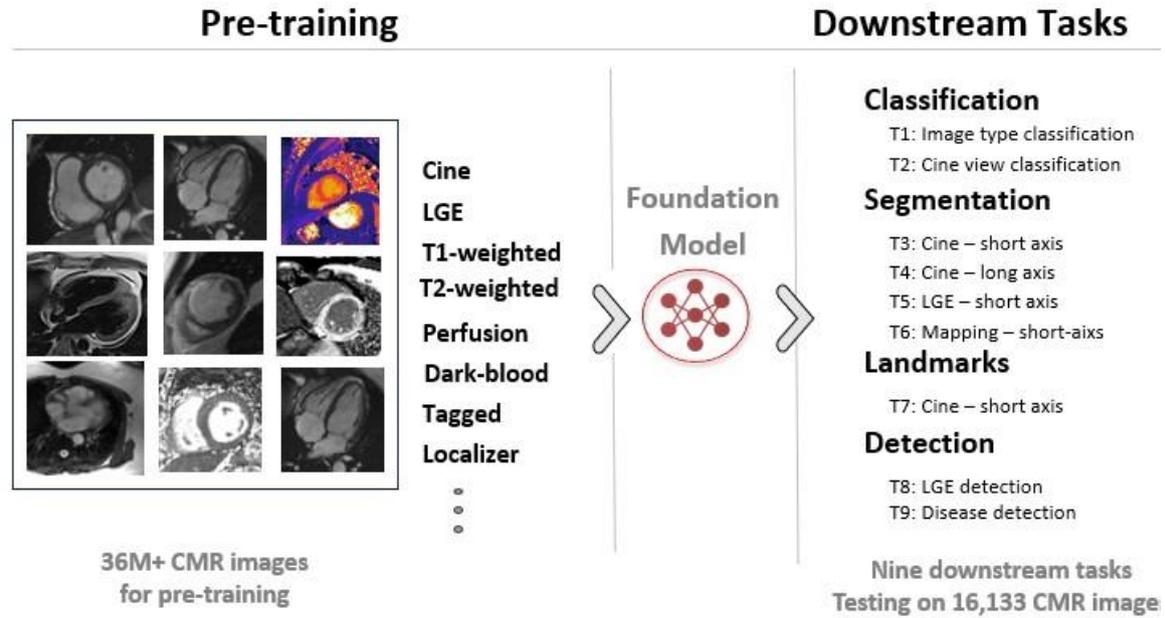

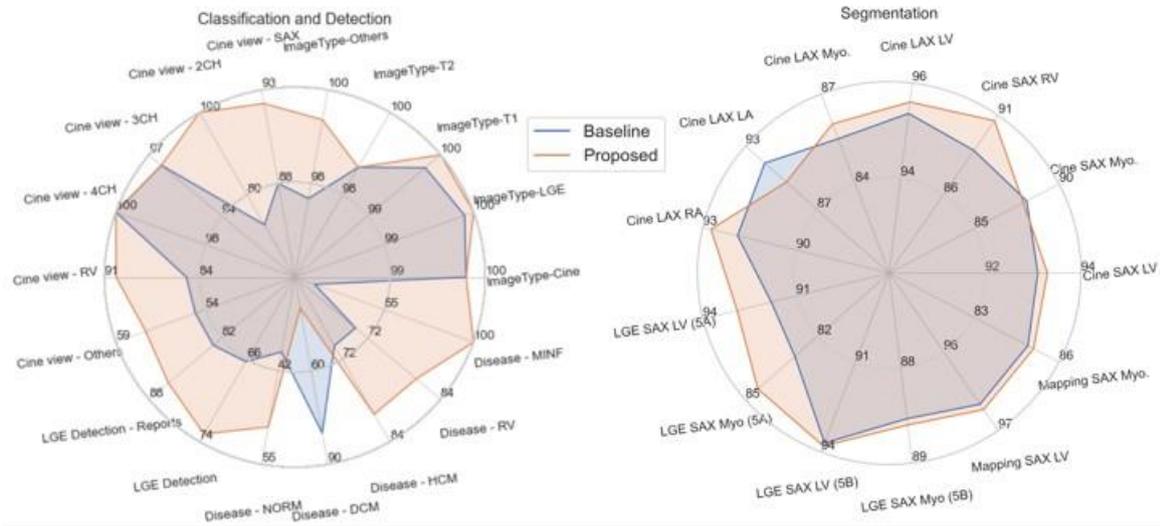

Fig. 1. Proposed method. A ViT-S model is pretrained on 36 million cardiac CMR images from 27,000 patients. The model was then finetuned on 11 downstream tasks typical to a CMR assessment workflow.

On the other hand, foundational models (FM) represent the latest generation of AI models that promise generalizability across data distributions and across different tasks. These are large models, trained on massive, diverse datasets. As a result, they have achieved SoTA results and excellent zero-shot and few-shot performance and generalizability in natural image processing(Oquab et al., 2024; Kirillov et al., 2023). FM are often vision transformers, which by being able to better capture global relationships in images, can overcome certain shortcomings of fully convolutional networks(Willemink et al., 2022). However, they require very large amounts of training data to reach their full potential. Thus, in the field of medical image analysis, task-specific models are still mainly used, especially for obtaining clinical metrics or disease diagnoses. While generalist FM show impressive transferability to medical domains (Baharoon et al., 2023), they still often lag behind fully supervised training in terms of task specific accuracies (Zhang and Metaxas, 2023). Compared to natural images, medical images offer different, but significant challenges due to its diverse modalities, contrast properties, different dimensionality and reconstruction methods which can lead to subtle and fine-grained differences between images, varying resolutions, and the long tail distribution of abnormalities. Most medical vision tasks are also low-data tasks, with few labeled examples, due to the high cost (expertise, time etc.) required to create annotations. We hypothesize that a modality-specific or organ-specific FM could provide stronger image representations relevant to the downstream tasks. This could in turn lead to better accuracies and robustness for related clinical tasks, reduce the need for labeled data, and offer an efficient path to model deployment for new tasks.

In this study, we propose a foundation model trained in a task-agnostic way on 36 million CMR images, the first such FM for CMR to our knowledge. We show that this model can be finetuned for classification, segmentation, regression, and detection tasks typical to a CMR image analyses workflow and demonstrate performance improvements relative to a comparable model, with no other changes in the pipeline. We also conduct ablation studies to demonstrate the importance of pretraining on relevant medical images, as opposed to natural images. In addition, we demonstrate the data efficiency of such a targeted FM by exploring its few-shot learning capabilities.

## 2. Related Works

### 2.1. Cardiac MR imaging tasks

Many studies explore CMR imaging tasks separately. Chauhan et al. (2022) report an accuracy of 0.90-0.95 for distinguishing between short-axis and 3 long-axis views. The study trained a convolutional network on a dataset of 2000 images. The dataset was actively augmented with views of complex anatomy and balanced across classes to achieve the best results. In the case of segmentation, studies report best Dice scores of 0.90 to 0.95 for LV(Bernard et al., 2018; Suinesiaputra et al., 2022; Schilling et al., 2024; Bai et al., 2018), 0.90-0.92 for RV(Bernard et al., 2018; Schilling et al., 2024; Åkesson et al., 2023) and 0.88 to 0.91 for myocardium(Bernard et al., 2018; Suinesiaputra et al., 2022; Schilling et al., 2024; Bai et al., 2018) in short axis (SAX) cine MRI. Many of these studies were done on public datasets, made available as part of challenges. "Automatic Cardiac Diagnosis Challenge" dataset (ACDC)(Bernard et al., 2018) is one such publicly available and fully annotated dataset for the purpose of (SAX, cine) MRI assessment. It consists of 150 patients with ground truth for LV, myocardium and RV segmentations in ED and ES, as well as an expert-assigned disease label. Isensee et al. achieved the overall top-performing method in the challenge by using an ensemble of 2D and 3D U-Net architectures to perform the segmentation. A random forest classifier was then used on top of clinical metrics handcrafted from the segmentations to predict disease classes. In long axis (LAX) views, Bai et al. (2018) trained a DL network on a large scale dataset of 4875 subjects from the UKBB cohort (Sudlow et al., 2015) and reported dice scores 0.94 for LV, 0.88 for myocardium, 0.95 for LA and 0.96 for RA. In the case of LGE images, some studies report dice scores of 0.84 to 0.88 for the short-axis myocardium(Zhuang et al., 2022; Zhang, 2021). EMIDEC(Lalande et al., 2020) dataset provides LGE images and myocardial contours for 100 patients.

The best performing method (Zhang, 2021) uses a cascaded 2D and 3D U-Net for segmenting the myocardium. Similarly, a few studies report dice scores of 0.85 to 0.86(Kalapos et al., 2023; Fahmy et al., 2019) for the myocardium on T1 and T2 mapping images. Kalapos et al. (2023) trained a U-Net architecture with ResNet5O encoder on 7000 T1 and T2 maps of differing etiologies, with additional contour-based post-processing to obtain an average myocardial dice score of 0.86. Other public datasets include the 2015 Kaggle Second Annual Data Science Bowl (Newton et al., 2015), which provided cine images for 700 patients, with the aim of detecting ED and ES volumes. Ground truth ED and ES volumes were provided, but no manually segmented contours. MyoPS dataset (Li et al., 2023) provides paired cine, LGE, and T2 scans for 45 patients along with ground truth contours for myocardium, scar, and edema. We refer the reader to this review(El-Taraboulsi et al., 2023) for further details. The public datasets, while publicly available, are on very small cohorts, with a limited number of centers, or are limited to cine MRI.

### 2.2. Foundation Models

Foundation models (FM) can be supervised (such as CLIP,Radford et al. (2021)), weakly supervised (SAMKirillov et al. (2023)) or self-supervised (DINO,Caron et al. (2021)). Self-supervised methods provide a promising option in medical imaging where labeled data is scarce. Baharoon et al. (2023) explored DINOv2(Oquab et al., 2024) pretrained models on classification and segmentation in radiology tasks, and report promising results. However, performance varies depending on the complexity of the task, the anatomy involved, the amount of training data available, and the finetuning method used. Ghesu et al. (2022) train a medical FM with 100 million images, and report a significant increase across multiple detection tasks in chest radiographs and brain MRI. Tang et al. pretrained a UNetr model 3D computed tomography images (CT) and obtained SoTA results on various segmentation tasks in CT images. Lu et al. (2024) trained a vision-language FM on 1.17M histopathological images, and showed SoTA results on various related downstream tasks with or without finetuning.

## 3. Methods

### 3.1. Foundation model for cardiac MR

We pretrain a ViT-S/8 (Touvron et al., 2021) model in a selfsupervised manner using DINO(Caron et al., 2021) on 36 million cardiac MRI images from 27,524 subjects from 3 sources: two clinical centers (henceforth referred to as Centers 1 and 2) and the UK Biobank (UKBB)(Sudlow et al., 2015) cohort. Datasets from Centers 1 and 2 are composed of comprehensive CMR studies including cine, LGE, T1 and T2 weighted imaging, flow imaging, black blood sequences, localizers, etc., from both healthy and diseased subjects. More information is given in Appendix 6.1. The images focus

4                                      *Jacob et al.*Table 1. Summary of all downstream tasks. Data numbers are given as "#images (#patients)". Tasks marked with * are in-distribution, with all other being out-of-distribution. Abbreviations: Myo: myocardium; Anth: Antherior; Inf: Inferior; Others, as defined in Fig. 2. Description of data from Centers 1-3 is given in the Appendix.

| Task No. | Task | Dataset | | | | |
|---|---|---|---|---|---|---|
| | | Source | #Classes | Train | Validation | Test |
| Classification | | | | | | |
| 1 | Image type classification | Center 3 | 5 (Cine, LGE, T1, T2, Others) | 24890 (100) | 4310 (15) | 8984 (29) |
| 2 | Cine view classification | Center 3 | 6 (SAX, 2CH, 3CH, 4CH, Aorta, Others) | 1457 (165) | 171 (16) | 4494 (165) |
| Segmentation | | | | | | |
| 3 | Cine SAX | ACDC (Bernard et al. (2018)) | 3 (LV, Myo, RV) | 150 (75) | 50 (25) | 100 (50) |
| 4 | Cine LAX (4CH) | Kaggle (Newton et al., 2015) | 4 (LV, Myo., RA, LA) | 346 (173) | 70 (35) | 458 (228) |
| 5A | LGE SAX | EMIDEC (Lalande et al., 2020) | 2 (LV, Myo.) | 70 (70) | 15 (15) | 15 (15) |
| 5B* | LGE SAX | EMIDEC, Center 1, Center 2 | 2 (LV, Myo.) | 6105 (808) | 1280 (176) | 1197 (159) |
| 6 | Mapping SAX | Center 3 | 2 (LV, Myo.) | 877 (100) | 128 (15) | 261 (29) |
| Landmarks localization | | | | | | |
| 7* | Cine SAX | UKBB (Sudlow et al., 2015) | 2 (Anth., Inf. RVIP) | 3198 (716) | 398 (80) | 399 (80) |
| Detection | | | | | | |
| 8A* | LGE – Clinical Reports | Center 2 | 2 (LGE/None) | 694 (347) | 158 (79) | 110 (55) |
| 8B | LGE Detection | EMIDEC | 2 (LGE/None) | 70 (70) | 15 (15) | 15 (15) |
| 9 | Disease detection | ACDC | 5 (NORM, DCM, HCM, RV, MINF) | 150 (75) | 50 (25) | 100 (50) |

on cardiac views and anatomies, with a large proportion being short-axis images, followed by long-axis (2, 3, and 4 chambers). Since the region of interest in a typical DICOM image from a cardiac study is limited, relative to the entire image size, all images were preprocessed by rescaling to a 1 mm x 1 mm resolution, and center-cropping to a size of 224x224. The model was trained for 7 days on 8 Nvidia Tesla H100 (80GB) GPUs, with a batch size of 1024 and patch size of 8 for the ViT.

*3.2. Downstream tasks*

We finetune and test the model separately for 9 tasks in classification, segmentation, landmark localization, and pathology detection, across cine, LGE, and mapping images, from multiple datasets. The tasks detailed below are illustrated in Figure 2.1. Data used for each task is summarized in Table 1. Public datasets and labels were used when available. In other cases, clinical data from three centers (Centers 1-3, 1 and 2 as in the previous section), were annotated by trained CMR experts. More information about the private datasets is given in Appendix 6.1. For each task, the available data was divided into training, validation, and testing split, on the patient level. Image type classification (task 1) aims to recognize different CMR acquisition protocols that lead to different tissue contrasts. Each image type can cover diverse views of the heart (short, long axis etc.). In this case, we classify between cine (a balanced steadystate free precession – bSSFP), T1 (a MOLLI acquisition for T1 pre- and post-contrast, including both the maps and the different T1-weighted images), T2 (Single-shot T2 prepared bSSFP, both the maps and the different T2-weighted images) and LGE (single-short or segmented PSIR acquisition). Cine view classification (task 2) is a fundamental task for automated CMR analyses, and attempts to choose relevant cine views for further analyses. While SAX and LAX views are typically studied, a CMR study can include many diverse views. This is represented in our study through the more uncommon Aorta focussed view and the "Others" class which includes other views as well as non-diagnostic images from the existing named classes. Tasks 1 and 2 use data from Center 3.

For CMR segmentation, in task 3 we segment the LV blood pool, myocardium and RV blood pool in SAX cine bSSFP images from the ACDC challenge. In task 4 we segment all four cardiac chambers and the myocardium in LAX 4-chamber view images from the Kaggle dataset. We present two experiments for the myocardium segmentation in LGE SAX images (tasks 5A and 5B). To explore the effect of the size of the datasets on the same task, we compare the segmentation performance on only the EMIDEC dataset for task 5A vs. adding more clinical data from Center 1 and Center 2 amounting to a 10-fold increase for task 5B. We note that this is an annotated subset of the larger dataset used from Centers 1 and 2 in the pretraining task, making task 5B an in-distribution downstream task. Mapping segmentation (task 6) includes T1 maps (pre- and post-contrast) and T2 maps. For task 7, landmark localization is performed on the UKBB dataset for the anterior and inferior right ventricular insertion points (ant. RVIP and inf. RVIP) in short-axis cine bSSFP images. While only the ant. RVIP is typically used clinically for AHA segment initialization, we also include the inf. RVIP. While tasks 8A and 8B both involve the detection of enhancement from LGE images, the GT (presence of LGE) in task 8A is extracted automatically from clinical reports, while task 8B used the publicly provided annotations (scar masks) to infer the same. Both tasks use 3 image slices extracted from the image stack at basal, mid, and apical levels as the input, and predict a binary label. Task 9 uses 3 frames from the cardiac cycle (ED, halfway between ED and ES, and ES), extracted at the basal level from the image stack as the input, and predicts the patient as normal (NORM), or assigns a disease label dilated cardiomyopathy (DCM), hypertrophic cardiomyopathy (HCM), abnormal RV (RV) and previous myocardial infarction (MINF).

Since the focus of the study was to evaluate the pretraining strategy, no further optimization was done in terms of incorporating multiple spatial or temporal information, though that could potentially further improve the performance. Out of the 9 tasks, 6 are out-of-distribution, using datasets that were not seen during pretraining. The remaining 3 tasks (5B, 7 and 8A) are in-distribution, where we use a small subset of the pretraining data with annotations. It is to be noted that GT annotations were not used in the pretraining stage.

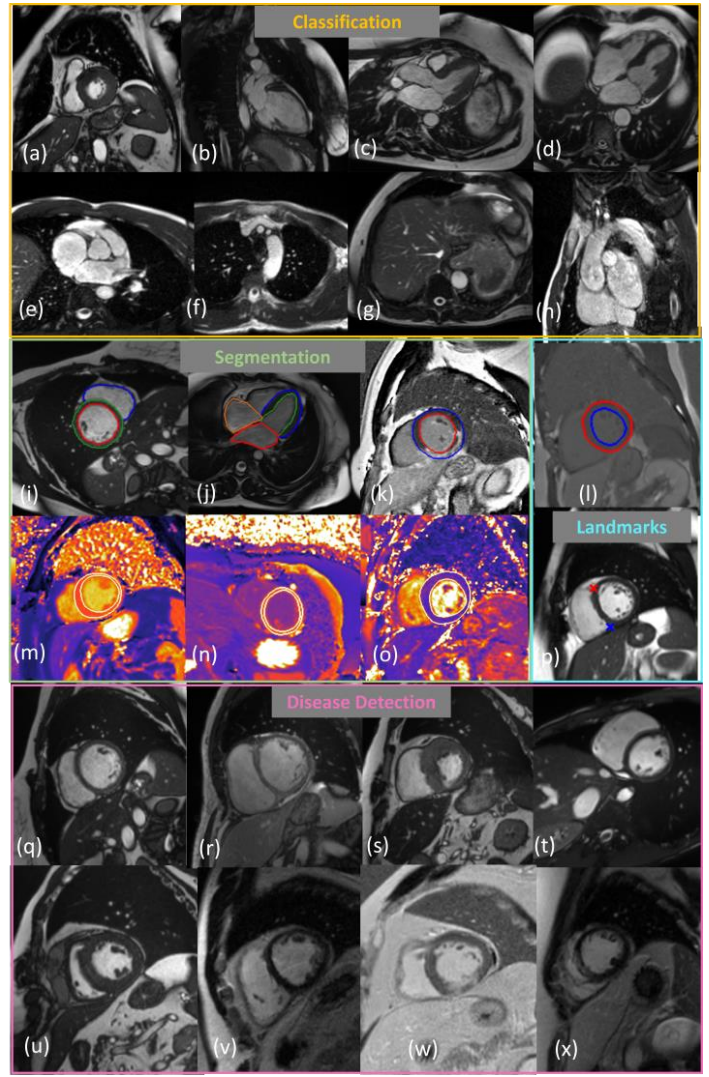

Fig. 2. Downstream tasks. Cine view classification: a) SAX, b) 2CH, c) 3CH, d) 4CH, e-f) Aorta, g-h) Other, i) Cine SAX segmentation, j) Cine LAX 4CH seg-mentation, k-l) LGE SAX segmentation, m-o) Mapping segmentation (Pre-contrast T1, post-contrast T1, T2, respectively) , p) Anterior and Inferior RVIP localization on cine SAX, q-u) Disease detection (NORM, DCM, HCM, abnormal RV, MINF respectively), v-x) LGE detection (No LGE, LGE present, LGE present, respectively). Abbreviations: SAX: short axis; LAX: long axis, CH: chamber, RVIP: RV insertion point; NORM: Normal; DCM: Dilated Cardiomyopathy; HCM: Hypertrophic cardiomyopathy; MINF: myocardial infarction.

### 3.3. Supervised fine-tuning

For every task, we consider two experiments: a) Baseline: ResNet50 (He et al., 2016)initialized with ImageNet pretrained weights (23M parameters) b) The ViT-S encoder pre-trained on cardiac MR images (21M parameters). ResNet50 was chosen for its similar number of parameters. Both networks were trained on labeled data available for each task, in an identical training setup.

For classification and detection tasks, a single linear layer was added on top of the features from the respective encoders, similar to the linear evaluation in this study(Caron et al., 2021). Freezing/finetuning of the encoder was treated as a hyperparameter, and chosen based on performance on the respective validation split for each experiment. For the ResNet50 encoder, the feature vector was extracted and averaged along the spatial dimension to get a feature vector of dimension 2048. For the ViT-S, we concatenate the [CLS] token of the last 4 blocks, along with the mean of the patch tokens from the last block, to get a feature vector of dimension 1920. Both models were trained with Cross Entropy loss.

For segmentation and landmark tasks, the trained ResNet50 and ViT-S/8 encoders were used with UNet decoder to create a ResNet50-Unet(Iakubovskii, 2019) or a UNETR(Hatamizadeh et al., 2021) respectively. The models were trained with Jaccard loss(Bertels et al., 2019), with Softmax activation on the last layer. Prior studies(Baharoon et al., 2023) as well as our empirical experiments demonstrated higher accuracies for end-to-end finetuning, so both models were finetuned end-to-end for the segmentation tasks. The landmark localization was treated as a regression task, with the network trained to predict a Gaussian heatmap centered on the landmark, similar to the set up in Xue et al. (2021).

In all tasks (except for the detection tasks, as noted in Section 3.2) the model works on a single image, repeated in the channel dimension to create a 3-channel input. All images were resampled to 1 mm x 1 mm resolution, and center cropped to 224 x 224. Intensities were capped at 98 percentile and then normalized to -1 to 1. For each task, the available data was divided into training, validation, and testing split, on the patient level. The best model was chosen based on the validation metric, averaged over classes, on the specific validation split. The metrics chosen were the Dice coefficient for segmentation, per-class accuracy for classification (number of correct predictions/total number of samples per class) and detection tasks, and the absolute Euclidean distance error in mm for landmark tasks.

### 3.4. Effect of pretraining on CMR data

An alternative to pretraining on CMR data is to pretrain the FM on natural images. To evaluate the benefit of training specifically on CMR data, we obtain the weights of a ViT-S/8 pretrained on natural images(Caron et al., 2021), and finetune it in the same framework. We choose three representative tasks for the comparison: cine view classification (task 2), LGE SAX segmentation (task 5A) and mapping SAX segmentation (task 6).

### 3.5. Few-shot performance

To simulate a low annotated data task often encountered in medical image analysis settings, we evaluated the few-shot performance of both the baseline and the proposed model. The amount of training data was varied systematically between 1512 samples per class, as available, in two representative tasks – cine view classification (task 1), and mapping SAX segmentation (task 6). The encoder was kept frozen, to prevent overfitting.

## 4. Results

### 4.1. Downstream tasks

The results are shown in Table 2. The proposed model outperforms the baseline model in the majority of the cases. Both models perform well in distinguishing different CMR image types, with minor improvements from the SSL pretrained model, especially for the "Others" class. For the cine view classification task (task 2), we observe an increase of 6.8 percentage points (pp) over all classes, ranging from 4.7 pp in SAX to 27.2 pp for the "2CH" class. While the proposed model achieves excellent accuracies for the more common views (short and long-axis), both models face challenges with the "Others" class, which is a heterogenous class, consisting of other views (such as LVOT, 5CH, etc.) as well as non-diagnostic images from regular views (such as chamber not visible). For the segmentation tasks, we obtain improvements from 0.1 to 1.8 pp, depending on the class and task. It is interesting to note that we get greater improvements in task 5A, relative to task 5B for LGE SAX segmentation. This could be related to the smaller finetuning dataset in task 5A, where the learned features of the CMR SSL pre-trained model provides a greater benefit. For the landmark detection task, we achieve mixed results, with the proposed model outperforming the baseline for one landmark, and vice versa for the other. In both segmentation and landmark detection, we also observe generally lower standard deviations in the metrics for the proposed method. In the pathology detection tasks, the proposed model obtains higher average accuracies in all tasks. We observed improvements of 3.7 pp and 6.6 pp for the LGE detection tasks, with the greater improvement in task 8B, with the smaller amount of data. The results from the CMR-SSL model are presented against performance reported in prior studies (where available) in Table 3. Performance from prior studies is presented as a range due to the fragmented nature of the field, with evaluations on private datasets being the norm. The out-of-box performance of the proposed model (fine-tuned without any task-specific optimization of method or hyperparameters), is comparable or superior in many tasks across classification, segmentation, and landmark localization.

For the disease detection (task 11), the proposed model outperforms the baseline by 14 pp. Recognizing cardiomyopathies is a complex task, and more careful study design such as incorporating multiple spatial locations and temporal dynamics through metrics like ejection fraction will almost certainly improve performance for both models, similar to the SoTA (Table 3). However, the scope of the current study is to assess the information content of the raw features obtained from both the encoders. We see that the SSL pretrained model extracts more relevant features, as evidenced by the higher "out-of-box" accuracy when compared to the baseline method.

Table 2. Results for the baseline and proposed model across all tasks. The best result for each task is highlighted in bold. For classification, segmentation, and detection tasks, metrics range from 0 to 1, with the higher the better. For landmark localization, metric represents distance error in mm, with the lower the better. Abbreviations are as defined in Fig. 2. and Table

### Classification

| Task No. | Task | Class | Accuracy(0-1) Baseline | Proposed |
|---|---|---|---|---|
| 1 | Image type classification | Total | 0.994 | **0.998** |
|  |  | Cine | **0.998** | **0.998** |
|  |  | LGE | 0.999 | **1.000** |
|  |  | T1 | 0.998 | **1.000** |
|  |  | T2 | **0.990** | **0.990** |
|  |  | Others | 0.971 | **0.992** |
| 2 | Cine view classification | Total | 0.827 | **0.896** |
|  |  | SAX | 0.875 | **0.922** |
|  |  | 2CH | 0.727 | **1.000** |
|  |  | 3CH | **0.964** | **0.964** |
|  |  | 4CH | **1.000** | **1.000** |
|  |  | Aorta | 0.854 | **0.902** |
|  |  | Others | 0.550 | **0.575** |

### Segmentation

| Task No. | Task | Class | Dice score (0-1) Baseline | Proposed |
|---|---|---|---|---|
| 3 | Cine SAX | LV | 0.931 (0.17) | **0.933 (0.14)** |
|  |  | Myo. | **0.881 (0.16)** | 0.879 (0.14) |
|  |  | RV | 0.890 (0.20) | **0.907 (0.18)** |
| 4 | Cine LAX (A4C) | LV | 0.952 (0.06) | **0.955 (0.03)** |
|  |  | Myo. | 0.855 (0.07) | **0.860 (0.06)** |
|  |  | LA | **0.914 (0.11)** | 0.896 (0.12) |
|  |  | RA | 0.917 (0.13) | **0.927 (0.06)** |
| 5A | LGE SAX | LV | 0.918 (0.10) | **0.929 (0.05)** |
|  |  | Myo. | 0.826 (0.12) | **0.844 (0.07)** |
| 5B | LGE SAX | LV | 0.937 (0.05) | **0.938 (0.05)** |
|  |  | Myo. | 0.883 (0.07) | **0.884 (0.06)** |
| 6 | Mapping SAX | LV | 0.965 (0.02) | **0.966 (0.02)** |
|  |  | Myo. | 0.849 (0.08) | **0.851 (0.07)** |

### Landmark localization

| Task No. | Task Name | Distance error (mm) Baseline | Proposed |
|---|---|---|---|
| 7 | Cine SAX Anth. RVIP | 2.013 (1.7) | **1.914 (1.6)** |
|  | Cine SAX Inf. RVIP | **1.679 (1.3)** | 1.774 (1.3) |

### Detection

| Task No. | Task | Class | Accuracy (0-1) Baseline | Proposed |
|---|---|---|---|---|
| 8A | LGE Detection (Reports) | Total | 0.827 | **0.864** |
| 8B | LGE Detection | Total | 0.667 | **0.733** |
| 9 | Disease Detection | Total | 0.560 | **0.700** |
|  |  | NORM | 0.400 | **0.500** |
|  |  | DCM | **0.800** | 0.400 |
|  |  | HCM | 0.700 | **0.800** |
|  |  | RV | 0.700 | **0.800** |
|  |  | MINF | 0.200 | **1.000** |

Table 3. Comparison of the results with prior literature, where available. Rows where the CMR-SSL model is within or better than the reported range are marked in bold. For classification, segmentation and detection tasks, metrics range from 0 to 1, with the higher the better. For landmark localization, the lower the better.

### Classification

| Task No. | Task | Class | Accuracy(0-1) | |
|---|---|---|---|---|
| | | | Prior Studies | Proposed |
| 2 | Cine view classification | Average (Only SAX, 2CH, 3CH, 4CH) | 0.90-0.95 (Chauhan et al., 2022) | 0.972 |

### Segmentation

| Task No. | Task | Class | Dice score (0-1) | |
|---|---|---|---|---|
| | | | Prior Studies | Proposed |
| 3 | Cine SAX | LV | 0.90-0.95 (Bernard et al., 2018; Suinesiaputra et al., 2022; Schilling et al., 2024; Bai et al., 2018) | 0.933 |
| | | Myo. | 0.86-0.91 (Bernard et al., 2018; Suinesiaputra et al., 2022; Schilling et al., 2024; Bai et al., 2018; Penso et al.) | 0.879 |
| | | RV | 0.90-0.92 (Bernard et al., 2018; Suinesiaputra et al., 2022; Schilling et al., 2024; Bai et al., 2018) | 0.907 |
| | Cine LAX 4 (A4C) | LV | 0.89-0.94 (Bai et al., 2018; Shahzad et al.) | 0.955 |
| | | Myo. | 0.88(Bai et al., 2018) | 0.860 |
| | | LA | 0.90-0.95 (Bai et al., 2018; Zhang et al.) | 0.896 |
| | | RA | 0.95-0.96 (Bai et al., 2018; Regehr et al.) | 0.927 |
| 5B | LGE SAX | Myo. | 0.84-0.88 (Zhuang et al., 2022; Zhang, 2021) | 0.884 |
| 6 | Mapping SAX | Myo. | 0.85-0.86 (Fahmy et al., 2019; Kalapos et al., 2023) | 0.851 |

### Landmark localization

| Task No. | Task | Name | Distance error (mm) | |
|---|---|---|---|---|
| | | | Prior Studies | Proposed |
| 7 | Cine SAX | Anth. RVIP | 3.1 (Xue et al., 2021; Ghadimi et al.) | 1.914 |

### Detection

| Task No. | Task | Class | Accuracy (0-1) | |
|---|---|---|---|---|
| | | | Prior Studies | Proposed |
| 9 | Disease Detection | Total | 0.86-0.96 (Bernard et al., 2018) | 0.700 |

### 4.2. Effect of pretraining on CMR data

In most settings (Table 4), the CMR pretrained model outperforms the ViT-S/8 model pretrained on natural images (NI-ViTS). Interestingly, the NI-ViTS achieves competitive performance in the classification task on some classes, while lagging behind both the proposed and the baseline training in segmentation tasks, with a wider gap in the task with the smaller finetuning dataset (task 5A). This trend is also observed in the cardiac segmentation task in this study(Baharoon et al., 2023). This might be explained by the fact that classification can be done on more global features while segmentation requires fine-grained features at a smaller scale. Pre-training on CMR images provides more relevant representations at the finer scale.

Table 4. Results on 3 representative tasks for the baseline, the proposed CMR pretrained model, and natural image pretrained model. The best result for each task is highlighted in bold, and the second-best result is underlined, when applicable. Dice scores are expressed as means and standard deviations.

| Task No. | Task | Class | Baseline | Proposed-CMR pretrained | NI Pretrained |
|---|---|---|---|---|---|
| Classification - Accuracy | | | | | |
| 2 | Cine view | Total | 0.827 | **0.896** | 0.713 |
| | | SAX | 0.875 | **0.922** | 0.813 |
| | | 2CH | 0.727 | **1.00** | 0.705 |
| | | 3CH | **0.964** | **0.964** | 0.929 |
| | | 4CH | **1.00** | **1.00** | **1.00** |
| | | Aorta | 0.854 | **0.902** | 0.195 |
| | | Others | 0.550 | 0.575 | **0.600** |
| Segmentation – Dice Score | | | | | |
| 5A | LGE SAX | LV | 0.918 (0.10) | **0.929** (0.05) | 0.879 (0.14) |
| | | Myo. | 0.826 (0.12) | **0.844** (0.07) | 0.753 (0.14) |
| 6 | Mapping SAX | LV | 0.965 (0.02) | **0.966** (0.02) | 0.959 (0.02) |
| | | Myo. | 0.918 (0.10) | **0.929** (0.05) | 0.879 (0.14) |

*4.3. Few-shot performance*

Figure 5 shows the test metric averaged across classes for cine view classification, and Mapping segmentation when using very few labeled samples for training. The proposed model outperforms the baseline model in all configurations. It also reaches convergence accuracy faster than the baseline model.

5. Conclusions

In this study, we evaluated a vision foundation model trained in a self-supervised manner, on large amounts of CMR data on a wide variety of downstream tasks relevant to a clinical workflow. We achieve performance comparable to the stateof-the-art for classification, segmentation, and landmark localization tasks, and promising results for disease detection, with no task-specific optimization. We compare against a comparable fully supervised DL network, without large-scale pretraining and demonstrate that targeted SSL pretraining can benefit all tasks in terms of accuracy and robustness, across a wide range of resource (labeled data) settings. The proposed method thus presents a resource-efficient, unified framework to tackle a cardiac MR imaging workflow, providing opportunities for faster time to deployment in real world settings. While pretraining the FM is a computationally intensive exercise, it is a onetime activity, providing an encoder that can then be used across different tasks. This study explored the feasibility of such a framework in very basic settings. Further task-specific optimization, motivated by clinical knowledge of the task can potentially improve the results, especially in more complex tasks such as disease detection. Parameter efficient training methods such as LoRA(Hu et al., 2021) could provide faster pretraining times. Model distillation methods can be used to obtain a more lightweight model for deployment. Another avenue of research is to create a unified, cardiac-specific FM trained on images from multiple modalities such as cardiac computed tomography and echocardiography. Such a model could implicitly utilize the strengths of various imaging modalities to provide stronger image representations. In addition, weakly supervised training can enable learning from accompanying clinical reports to create multi-modal foundation models.

Acknowledgements: This research was conducted using the UK Biobank Resource under application number 30769.

Disclaimer: The concepts and information presented in this paper are based on research results that are not commercially available. Future commercial availability cannot be guaranteed.

6. Appendix

*6.1. Private datasets*

Center 1: Data from this center constitutes of CMR studies performed on 1.5T scanners (MAGNETOM Aera, Siemens Healthineers AG, Erlangen, Germany). Long-axis and short axis views covering the entire LV were obtained using balanced steady-state free-precession sequence (b-SSFP). LGE images were acquired after injection of a bolus of gadolinium-based contrast agent (Dotarem, Guerbet, France, 0.1 mmol/kg). Stress perfusion imaging was performed using a saturation-prepared b-SSFP sequence. A series of six slices (four short-axis views, in addition to 2- and 4-chamber views) were acquired every other heartbeat. Single-breath-hold 3D T1-weighted inversion recovery gradient-echo sequence was acquired with the same prescriptions to detect LGE. The inversion time was individually adjusted to null normal myocardium. Other images in the dataset include those acquired with T2 weighted sequences such as HASTE and STIR, T2 star weighted, compressed sensing techniques, localizer scans, etc.

Center 2: The datasets from this center were acquired on 1.5T and 3T MRI systems (MAGNETOM Avanto and Skyra, Siemens Healthineers AG, Erlangen, Germany). CMR studies included cine bSSFP images in 2-, 3-, and 4-chamber LAX, and a stack of SAX slices. The protocol also included cine images focussed on the LV outflow track and aorta, at various planes. LGE images, both single-shot and segmented, were acquired as single-breath-hold T1-weighted inversion-recovery gradientecho images acquired 10 minutes after contrast injection. The dataset also included first-pass myocardial perfusion imaging at rest and adenosine stress. The patient cohort includes both normal patients, as well as patients with cardiomyopathies such as dilated cardiomyopathy, hypertrophic cardiomyopathy, ischemic heart disease, and myocarditis.

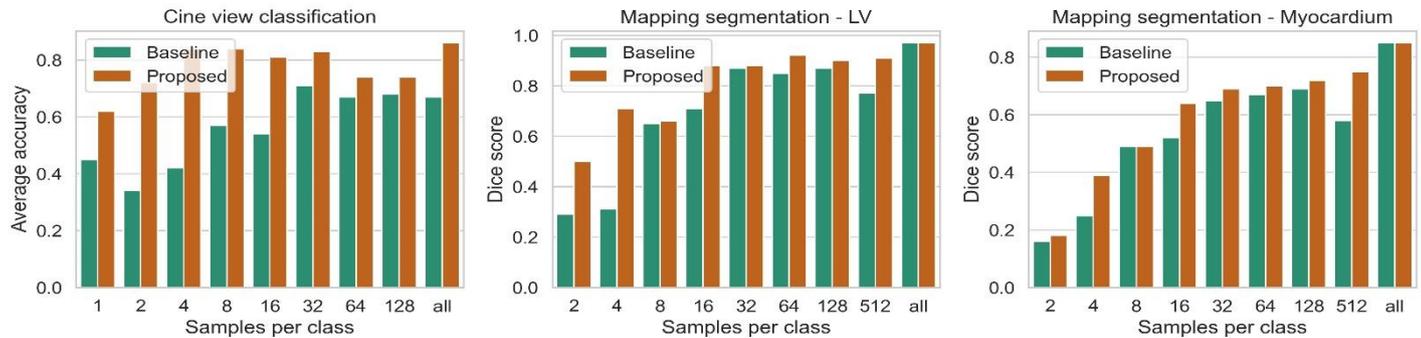

Fig. 3. Results of few-shot training for cine view classification and Mapping segmentation: Performance of the baseline and proposed model are shown in dotted and solid lines respectively.

Center 3: The data from this center includes 144 clinical subjects (52 normal, 49 myocarditis, 20 sarcoidosis, 23 systemic disease) were scanned on a 1.5T MRI system (MAGNETOM Aera, Siemens Healthineers AG, Erlangen, Germany). CMR studies included cine bSSFP images in 2-, 3-, and 4chamber LAX, and a stack of SAX slices. The cine imaging included views of the LV outflow tract and aorta. LGE imaging consisted of Inversion Recovery FLASH acquired 10 minutes after contrast injection in several short-axis planes. Native
and post-contrast T1 Modified Look-Locker Inversion recovery (MOLLI) and T2 prepared fast-low-angle shot maps were acquired. This dataset was not used in the pretraining stage.